\documentclass[jap,graphicx, reprint,unsortedaddress]{revtex4-1} 
\usepackage{amsfonts, amsmath, graphicx, feynmf}
\usepackage[version=3]{mhchem}

\begin{document} 
\title{Spin accumulation in asymmetric topological insulator thin films in out of plane magnetic fields} 
	\author{Zhuo Bin Siu}
	\affiliation{Computational Nanoelectronics and Nanodevices Laboratory,Electrical and Computer Engineering Department, National
		University of Singapore, Singapore}
	\email{elesiuz@nus.edu.sg }
	\author{Debashree Chowdhury}
	\affiliation{Department of Physics, ​Harish-Chandra Research institute 
		Chhatnag Road, Jhusi, Allahabad, 
		U. P. 211019,India}\email{debashreechowdhury@hri.res.in}
	\author{Banasri Basu}
	\affiliation{Physics and Applied Mathematics Unit, Indian Statistical Institute, Kolkata 700108, India}
	\email{sribbasu@gmail.com}
	\author{Mansoor B.A. Jalil}
	\affiliation{Computational Nanoelectronics and Nanodevices Laboratory,Electrical and Computer Engineering Department, National
		University of Singapore, Singapore}
	\email{elembaj@nus.edu.sg}

\begin{abstract} 
In this work we study the spin accumulation due to an in-plane electric field in an asymmetric topological insulator (TI) thin film system with an out of plane magnetic field and an in-plane magnetization. A TI thin film differs from the more typically studied thick TI system in that the former has both a top and a bottom surface where the states localized at both surfaces can couple to each other due to the finite thickness. In typical spin torque experiments on TI thin film systems, the top and bottom surfaces of the film are asymmetric as the former is in contact with a ferromagnetic layer while the latter is adjacent to a non magnetic substrate.  This may lead to differing (i) potentials and (ii) magnetization strengths experienced by the top and bottom surface states. We show, via Kubo formula calculations, that each of these two effects can lead to in-plane spin accumulation perpendicular to the magnetization direction which are otherwise absent in a top-bottom symmetric TI thin film system. This spin accumulation results from the breaking of the antisymmetry of the spin accumulation around the zero magnetic field equal energy contours.

\end{abstract} 
\maketitle
\section{Introduction}
In this work we calculate the spin accumulation due to an in-plane electric field in an asymmetric topological insulator (TI) thin film \cite{PRB80_205401,PRB81_041307,PRB81_115407}  with an in-plane magnetization and out of plane magnetic field. The spin accumulation results in a spin torque acting on the magnetization which may be used to switch the magnetization direction.  Whereas there have been numerous earlier works on TI thin films in out of plane magnetic fields \cite{SciRep5_13277,PRB89_155419,JPC26_165303,SciRep3_1261,JAP113_043720,PRB83_195413}, the inclusion of an in-plane magnetization has not yet been studied extensively. 

A TI thin film of finite thickness differs from a semi-infinitely thick TI slab in that the former possess both a top as well as a bottom surface where the surface states localized at each surface can couple to each other due to the finite thickness. The effective Hamiltonian of a TI thin film system with finite thickness along the $z$ direction subjected to a magnetization in the $x$ direction and an out of plane magnetic field in the $z$ direction may be written as 
\begin{equation}
	H_0 = v_f\tau_z(\vec{\pi}\times\hat{z}).\vec{\sigma} + \Delta_t \tau_x +  \Delta_z \sigma_z + M_x\sigma_x.  
\end{equation}

In the above, the $\pi_i$s are the gauge-invariant momenta $\pi_i \equiv (k_i + A_i)$ where $A_i$ is the $i$th component of the electromagnetic vector potential due to the $z$ magnetic field. (We set $e=\hbar=1$ throughout this paper.) The $\sigma$s are real spins and the $\tau$ correspond to whether the states are localized nearer the top ($\langle \tau_z \rangle = +1$) or bottom ($\langle \tau_z \rangle =-1$) surface. The $v_f\tau_z(\vec{\pi}\times\hat{z})\cdot\vec{\sigma}$ term hence corresponds to two copies of the the Dirac fermion Hamiltonian, one for the top surface and the other for the bottom one, the $\Delta_t \tau_x$ term the inter-surface coupling between the top and bottom surfaces, the $\Delta_z\sigma_z$ term the Zeeman energy due to the magnetic field and/or out of plane magnetization, and the $M_x\sigma_x$ the coupling with the in-plane magnetization.  

A typical experimental setup for studying spin torques in TI thin film systems consists of a TI thin film grown on top of a non-magnetic substrate and a ferromagnetic (FM) layer deposited on top of the FM layer. The magnetization of the FM layer couples to the spin accumulation in the TI  (for example, Refs. \onlinecite{PRL114_257202} and \onlinecite{Nat511_449} ). The asymmetry between the top and bottom surfaces of the TI surfaces can lead to two possible effects. First, the spin accumulation of the top TI surface may be more strongly coupled to the FM magnetization than the bottom layer due to the closer proximity of the latter. Second, there may be differing contact potentials at the top and bottom TI surfaces due to the different types of materials ( FM versus substrate) at the top and bottom surfaces of the thin film respectively. We model these two effects by the introduction of two terms to the Hamiltonian -- a $E_z\tau_z$ \cite{PRL111_146802} term corresponding to the potential difference between the top and bottom surfaces  , and a $\delta M_x \sigma_x \tau_z$ term for the differences in the coupling of the spin accumulation to the FM magnetization between the top and bottom layers. 

Introducing the ladder operators $a = \frac{l}{\sqrt{2}}(p_{x} - ip_{y}),~~a^{\dag} = \frac{l}{\sqrt{2}}(\pi_{x} + ip_{y}),$ with the magnetic length $l \equiv \sqrt{\frac{1}{B}}$, the full Hamiltonian reads
\begin{eqnarray}
	H = && \frac{i \omega}{\sqrt{2}} \tau_z (\sigma_+ a - \sigma_- a^\dag) \tau_z + \nonumber \\
	&&\Delta_t \tau_x +  \Delta_z \sigma_z + (M_x\mathbb{I}_\tau + \delta M_x\tau_z) \sigma_x + E_z \tau_z  \label{H1}
\end{eqnarray}
where $\omega \equiv  v_f/l$. 

In the \textit{absence} of the $(M_x\mathbb{I}_\tau + \delta M_x\tau_z)\sigma_x$ term the $z$ angular momentum is conserved, and $H$ can be rather easily diagonalized. Following Ref. \onlinecite{PRB83_195413} \footnote{It appears that there is a mistake in the corresponding Hamiltonian Eq. 5 there in which the $\omega_b$s should be replaced by $-\omega_b$. This mistake is carried forward into the expressions for the Landau level eigenstates there. }, we write $H \Big|_{M_x=\delta M_x=0}$ in the basis of the $|n,\sigma=\pm, \text{T}/\text{B}\rangle$ states where $n$ is the Landau level index, $\sigma=\pm$ is the spin $z$ up / down state, and T / B stands for \textit{T}op / \textit{B}ottom corresponding to $\langle \tau_z = \pm 1 \rangle$. In this basis, $H \Big|_{M_x=\delta M_x=0}$  can be broken up into uncoupled block matrices (written in the order of $|n-1, +z, \text{T}\rangle$, $|n-1, +z, \text{B}\rangle$,$|n, -z, \text{T}\rangle$, $|n,-z,\text{B}\rangle$) 

\[
	\begin{pmatrix} E_z+\Delta_z & \Delta_t & i\sqrt{2n}\omega & 0 \\
	\Delta_t & -E_z + \Delta_z& 0 & -i\sqrt{2n}\omega \\
	-i\sqrt{2n}\omega & 0 & E_z-\Delta_z & \Delta_t \\
	0 & i\sqrt{2n}\omega & \Delta_t & -E_z-\Delta_z \end{pmatrix}.
\]

This 4 by 4 matrix can be diagonalized in order to obtain the eigenspectrum for each value of $n$. In particular, when $E_z=0$, we have, for each value of integer $n \geq 0$ four eigenstates $|n, \alpha = 0, 1, s=\pm 1\rangle$ given by 
\begin{eqnarray}
&& |n\alpha s\rangle = \nonumber \\
&& |n-1,\uparrow,T\rangle (-i s (-1)^\alpha f_{n\alpha s+}) + |n-1,\uparrow, B \rangle (i (-1)^{\alpha+s} f_{n\alpha s+}) \nonumber \\
&&  + |n,\downarrow, T \rangle(-s f_{n\alpha s-})+|n,\downarrow,B\rangle f_{n\alpha s-}
\end{eqnarray}
where $f_{n \alpha s \pm} \equiv \frac{1}{2} \sqrt{1 \pm \frac{\Delta + s \Delta_t}{\epsilon_{n \alpha s}}}.$

The $n$ index in $|n,\alpha,s\rangle$ denotes the Landau level index of the constituent spin down states while $s=\pm 1$ gives the sign of expectation value of the $z$ angular momentum of the state.. The $\alpha=0$ ($\alpha=1$) states correspond to the particle (hole) states. 
Analytic expressions for the normalized eigenstates with \textit{finite} $E_z$ may also be obtained but these are rather more messy and not very informative and will not be stated explicitly. We shall nonetheless continue to denote the finite $E_z$ eigenstates as $|n,\alpha, s\rangle$. 

The inclusion of $M_x\sigma_x$ into the Hamiltonian breaks the $z$ angular momentum conservation. Analytic expressions for the eigenstates in terms of elementary functions can no longer be obtained. We shall instead include the effects of the magnetization terms $(M_x+\delta M_x\tau_z)\sigma_x$ perturbatively up to second order in $M_x$ and/or $\delta M_x\tau_z$. We verify the validity of our perturbative approximation by comparing the eigenspectrum obtained from the exact numerical diagonalization of Eq. \ref{H1}, and the second order perturbation expansion of the energy shift due to the magnetization terms, the difference between the energy spectrum when $M_x$ and $\delta M_x$ have finite values, and when both terms are zero)   
\[	
	\Delta E_{n\alpha s} = \sum_{n',\alpha',s' \neq n \alpha s} \frac{ |\langle n\alpha s| (M_x\sigma_x + \delta M_x\sigma_x\tau_z) | n'\alpha's'\rangle |^2  }{E_{n\alpha s} - E_{n'\alpha's'}}
\]

\begin{figure}[ht!]
	\centering
	\includegraphics[scale=0.5]{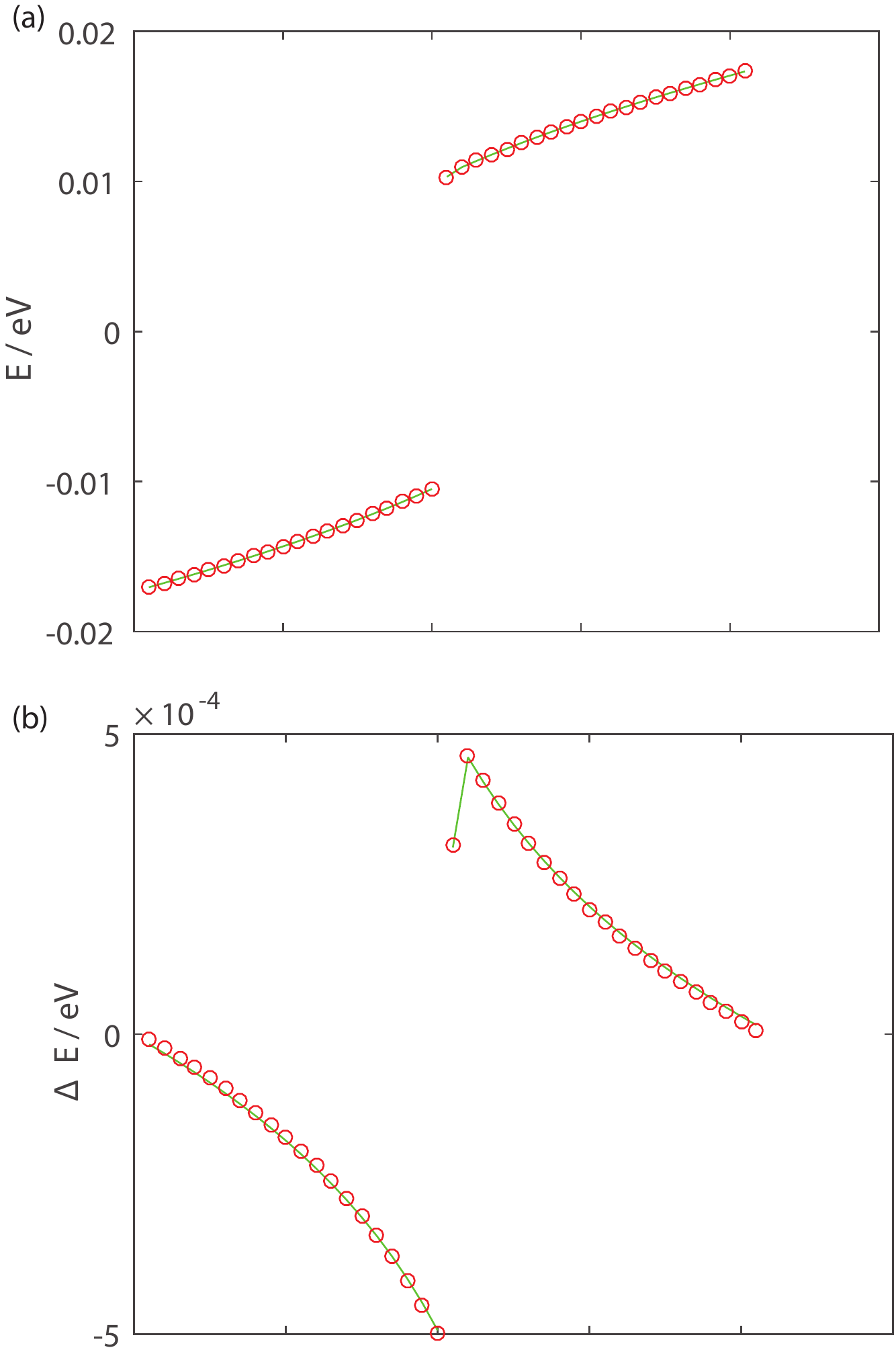}
	\caption{  Panel (a) shows the exact (continuous green line) and second order perturbative energy spectrum (red circles)  near zero energy arranged in ascending order of the Landau level energies for $\Delta_t = 10\ \mathrm{meV}$, $\Delta_z = 20\ \mathrm{meV}$, $M_x = 5\ \mathrm{meV}$, $E_z = 0.01\ \mathrm{meV}$ and $\delta M_x=0$ and $B_z = 30\ \mathrm{mT}$. Panel (b) shows the exact and second order perturbative energy shifts between $E_z = 0\ \mathrm{meV}$ and $E_z = 0.01\ \mathrm{meV}$ for the energy levels arranged in the same order as i n panel (a). }
	\label{gPerbComp} 
\end{figure}

Fig. \ref{gPerbComp} shows that the exact and perturbative values for the energy and energy shifts agree reasonably well for one exemplary set of parameters, which will be used for most of the numerical results that follow.

We note in passing that treating the magnetization terms perturbatively give good approximations to the exact eigenenergy spectrum only when the magnetization terms are weaker than the inter-surface coupling term $\Delta_t$. This is because a topological phase transition occurs when (for $\delta M_x=E_z=0$), $M_x > \Delta_t$ \cite{PRB83_195413}, which cannot be captured perturbatively.

\section{Kubo formula} 
We calculate the spin accumulation resulting from applying an in-plane electric field in the $i$th direction $E_i$ to the TI thin film by using the Kubo formula 
\begin{eqnarray}
&& \langle \delta O \rangle \nonumber  /  E_i = \\
&& \sum_{n\gamma \neq n'\gamma'} \mathrm{Im} \langle n \gamma^1|O|n'\gamma'^1 \rangle\langle n'\gamma'^1|J_i|n\gamma^1\rangle \frac{n_{n\gamma} - n_{n'\gamma'}}{(E_{n\gamma} - E_{n'\gamma'})^2} \label{kubo1} 
\end{eqnarray} 
where $O$ is an arbitrary operator, $J_i$  and $E_i$ are the $i$th component of the current and electric field respectively, and $\gamma$, $\gamma'$ are shorthand collective indices standing for $\alpha$ and $s$. The $1$ superscript in the bras and kets denote that these are the first order perturbed states $|n\gamma^1\rangle \equiv |n\gamma\rangle + |\delta n\gamma\rangle$ where $|n\gamma\rangle$ are the unperturbed eigenstates with $M_x=\delta M_x=0$, and $|\delta n\gamma\rangle$s the first order perturbed states \cite{JPD49_145003} given by the standard non-degnerate time-independent perturbation theory
\begin{equation}
	|\delta n\gamma^1\rangle = \sum_{n'\gamma' \neq n\gamma} |n'\gamma'\rangle \frac{\langle n'\gamma'|V|n\gamma\rangle}{E_{n\gamma}-E_{n'\gamma'}}
\end{equation}
where $V$ is either $M_x\sigma_x$ or $\delta M_x\sigma_x$. 
Some of the terms which give non-zero contributions up to second order in $|\delta n\gamma\rangle$ are (for notational simplicity we now lump all the state indices together and refer to them collectively as $a$,$b$,$c$, and $d$ ) 
\begin{eqnarray}
	&&(\mathrm{Im} \langle a|O|c\rangle\langle c|V_2|d\rangle\langle d|J|b\rangle\langle b|V_1|a\rangle)\times \nonumber \\
	 && \left( \frac{n_a-n_c}{(e_a-e_b) (e_c-e_a)^2 (e_c-e_d)} \right. \nonumber \\
	&&+\frac{n_c-n_b}{(e_a-e_b) (e_b-e_c)^2 (e_c-e_d)} \nonumber \\
	&&+\frac{n_b-n_d}{(e_a-e_b) (e_b-e_d)^2 (e_c-e_d)}  \nonumber \\ 
	&&+\left. \frac{n_d-n_a}{(e_a-e_b) (e_d-e_a)^2 (e_c-e_d)} \right).	\label{kub1}
\end{eqnarray} 

\begin{figure}[ht!]
	\centering
	\includegraphics[scale=1]{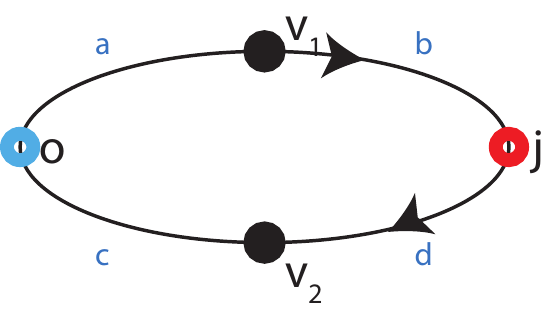}
	\caption{  The diagrammatic representations of the Kubo formula for an observable $O$ due to an in-plane electric field resulting in current $J$. The $V$ vertices represent either $M_x\sigma_x$ and / or $\delta M_x \sigma_x \tau_z$. } 
	\label{gKuboDiag} 
\end{figure}

These terms can be schematically represented in the Feynman diagram of Fig. \ref{gKuboDiag}. Each of the four terms corresponds to one of the four possible combinations of taking the difference between the Fermi-Dirac occupancy factors of one of the two upper lines and one of the two lower lines.

Indeed if one were to interpret the lines in Fig. \ref{gKuboDiag} as Matsubara Green's functions and read off from the diagram directly one would obtain 
\begin{eqnarray*}
	&& -\frac{1}{\beta} \sum_{iq_n} \Big( \mathcal{G}_a(i(\omega_n+q_n))O_{ac}\mathcal{G}_c(iq_n) \\ 
	&& V_{2;cd}\mathcal{G}_d(iq_n)J_{db}\mathcal{G}_b(i\omega_n+q_n)V_{1;ba} \Big).
\end{eqnarray*}  
Evaluating the Matsubara sum over $iq_n$, performing the analytic continuation $i\omega_n \rightarrow \omega + i\eta$ and then taking the limit $\omega \rightarrow 0$ and retaining only the non-divergent terms gives exactly the same terms as Eq. \ref{kub1}. 

It is instructive to first study the spin accumulation in the absence of asymmetry, i.e. when $\delta M_x = E_z = 0$. 
When $E_z=0$, we have
\begin{eqnarray}
	\langle n\alpha s|J_i|n'\alpha's'\rangle &\propto& \delta_{|n-n'|,1}\delta_{s,s'} \label{Ez0Ji} \\
	\langle n\alpha s|\sigma_i|n'\alpha's'\rangle &\propto& \delta_{|n-n'|, 1}\delta_{s,-s'} \label{Ez0s} 
\end{eqnarray} 
where $J_i$ and $\sigma_i$ are the current and spin operators in the $i$th direction, $i$ being on the $xy$ plane.

With these relations between the values of $n$ and the signs of $s$ in the `input' and `output' lines of the vertices in place, it is easy to see that there are no terms with up to two multiplicative factors of $M_x$ that will give a finite in-plane spin accumulation upon the application of an electric field. This is because in the zeroth order term, i.e. Fig. \ref{kub1} without the two $V$ vertices, the $O = \sigma_i$ vertex flips the signs of $s$ between its `input' and `output' lines but the $J_i$ vertex needs $s$ in both its input and output lines to have the same sign to give a finite contribution.  

The first order terms, i.e. Fig. \ref{kub1} without one of the $V$s on either the top or lower lines and with the remaining $V = M_x\sigma_x$, has zero contribution because of the mismatch between the $n$ indices. For example, one can see that after traversing the rest of the diagram the difference in the $n$ indices of the input and output lines of any of the three vertices ($J_i$, $V=M_x\sigma_x$ and $O=\sigma_j$) is 0 or $\pm 2$ whereas the difference has to be $\pm 1$ to yield a finite input. The second order term with $V_1=V_2=M_x\sigma_x$ also has zero contribution because after traversing the rest of the diagram the $s$ in the input and output legs of the $O=\sigma_i$ vertex have the same signs but we need them to have the same sign to obtain a finite contribution. 

The introduction of asymmetry breaks these restrictions, and leads to a finite $\langle \delta\sigma_i \rangle$ spin accumulation. We first discuss the effects of an asymmetric magnetization, modeled by the addition of a $\delta M_x\sigma_x$ term to the Hamiltonian. 

\section{Asymmetric magnetization}
When $E_z=0$, we have 
\[
	\langle n\alpha s|\tau_z\sigma_i|n'\alpha's'\rangle \propto \delta_{|n-n'|, 1}\delta_{s,s'}.
\]
Unlike a $\sigma_i$ vertex which only gives finite contribution for $s=-s'$,  the $\sigma_i\tau_z$ vertex has finite contributions for $s$ being of the same sign as $s'$. This leads to a diagram of the structure of Fig. \ref{kub1} with one of the two $V$ vertices being $M_x\sigma_x$ and the other being $\delta M_x\sigma_x\tau_z$ giving a finite contribution with $O=\sigma_i$. ( Diagrams with both $V_1=V_2=\delta M_x\sigma_x\tau_z$ do not give finite contributions. )  

On top of the summation over the four possible combinations of choosing one of the two upper lines and one of the two lower lines to take the difference of their Fermi-Dirac occupancy factors explicitly written out in Eq. \ref{kub1}, each of the four terms there usually gives six terms. These six terms correspond to the possible combinations of whether the $n$ indices are incremented or decreased by one across each of the four vertices $\sigma_i,V_1,V_2$ and $J_i$ while maintaining the requirement that the absolute value of the difference between the $n$ indices of the states at the input and output lines of every vertex remains one.  We also need to sum over the two choices of making $V_1=M_x\sigma_x$,$V_2=\delta M_x\sigma_x\tau_z$, and $V_1=\delta M_x\sigma_x\tau_z$,$V_2=\delta M_x\sigma_x$. (The only cases where there are less than six terms is when $n<0$ is encountered for some of the terms. )  Thus, suppressing the $\alpha$ state indices for notational simplicity and implicitly summing over the internal $\alpha$ indices, we have 

\begin{widetext}
\begin{eqnarray*}
&& \langle \delta \sigma_i \rangle = \sum_{n,s}\Big[ \\
&&\left(\frac{n_{n,s}-n_{-1 + n,s}}{(E_{-1 + n,s}-E_{n,s})^2}+\frac{n_{-1 + n,s}-n_{-1 + n,-s}}{(E_{-1 + n,s}-E_{-1 + n,-s})^2}+\frac{n_{n,s}-n_{n,s}}{(E_{n,s}-E_{n,s})^2}+\frac{n_{-1 + n,-s}-n_{n,s}}{(E_{n,s}-E_{-1 + n,-s})^2}\right)\times \\ 
&& \mathrm{Im} \left( \frac{\langle n,s|\delta M_x \sigma_x\tau_z|-1 + n,s\rangle\langle -1 + n,s|J_i|n,s\rangle\langle n,s|M_x\sigma_x|-1 + n,-s\rangle\langle -1 + n,-s|\sigma_i|n,s\rangle}{(E_{-1 + n,s}-E_{n,s}) (E_{n,s}-E_{-1 + n,-s})} \right) \\ 
&+&\left(\frac{n_{-2 + n,s}-n_{-1 + n,s}}{(E_{-1 + n,s}-E_{-2 + n,s})^2}+\frac{n_{-1 + n,s}-n_{-1 + n,-s}}{(E_{-1 + n,s}-E_{-1 + n,-s})^2}+\frac{n_{n,s}-n_{-2 + n,s}}{(E_{n,s}-E_{-2 + n,s})^2}+\frac{n_{-1 + n,-s}-n_{n,s}}{(E_{n,s}-E_{-1 + n,-s})^2}\right)\times \\ 
&& \mathrm{Im} \left( \frac{\langle n,s|\delta M_x \sigma_x\tau_z|-1 + n,s\rangle\langle -1 + n,s|J_i|-2 + n,s\rangle\langle -2 + n,s|M_x\sigma_x|-1 + n,-s\rangle\langle -1 + n,-s|\sigma_i|n,s\rangle}{(E_{-1 + n,s}-E_{n,s}) (E_{-2 + n,s}-E_{-1 + n,-s})} \right) \\ 
&+&\left(\frac{n_{n,s}-n_{-1 + n,s}}{(E_{-1 + n,s}-E_{n,s})^2}+\frac{n_{-1 + n,s}-n_{1 + n,-s}}{(E_{-1 + n,s}-E_{1 + n,-s})^2}+\frac{n_{n,s}-n_{n,s}}{(E_{n,s}-E_{n,s})^2}+\frac{n_{1 + n,-s}-n_{n,s}}{(E_{n,s}-E_{1 + n,-s})^2}\right)\times \\ 
&& \mathrm{Im} \left( \frac{\langle n,s|\delta M_x \sigma_x\tau_z|-1 + n,s\rangle\langle -1 + n,s|J_i|n,s\rangle\langle n,s|M_x\sigma_x|1 + n,-s\rangle\langle 1 + n,-s|\sigma_i|n,s\rangle}{(E_{-1 + n,s}-E_{n,s}) (E_{n,s}-E_{1 + n,-s})} \right) \\ 
&+&\left(\frac{n_{n,s}-n_{1 + n,s}}{(E_{1 + n,s}-E_{n,s})^2}+\frac{n_{1 + n,s}-n_{-1 + n,-s}}{(E_{1 + n,s}-E_{-1 + n,-s})^2}+\frac{n_{n,s}-n_{n,s}}{(E_{n,s}-E_{n,s})^2}+\frac{n_{-1 + n,-s}-n_{n,s}}{(E_{n,s}-E_{-1 + n,-s})^2}\right)\times \\ 
&& \mathrm{Im} \left( \frac{\langle n,s|\delta M_x \sigma_x\tau_z|1 + n,s\rangle\langle 1 + n,s|J_i|n,s\rangle\langle n,s|M_x\sigma_x|-1 + n,-s\rangle\langle -1 + n,-s|\sigma_i|n,s\rangle}{(E_{1 + n,s}-E_{n,s}) (E_{n,s}-E_{-1 + n,-s})} \right) \\ 
&+&\left(\frac{n_{n,s}-n_{1 + n,s}}{(E_{1 + n,s}-E_{n,s})^2}+\frac{n_{1 + n,s}-n_{1 + n,-s}}{(E_{1 + n,s}-E_{1 + n,-s})^2}+\frac{n_{n,s}-n_{n,s}}{(E_{n,s}-E_{n,s})^2}+\frac{n_{1 + n,-s}-n_{n,s}}{(E_{n,s}-E_{1 + n,-s})^2}\right)\times \\ 
&& \mathrm{Im} \left( \frac{\langle n,s|\delta M_x \sigma_x\tau_z|1 + n,s\rangle\langle 1 + n,s|J_i|n,s\rangle\langle n,s|M_x\sigma_x|1 + n,-s\rangle\langle 1 + n,-s|\sigma_i|n,s\rangle}{(E_{1 + n,s}-E_{n,s}) (E_{n,s}-E_{1 + n,-s})} \right) \\ 
&+&\left(\frac{n_{2 + n,s}-n_{1 + n,s}}{(E_{1 + n,s}-E_{2 + n,s})^2}+\frac{n_{1 + n,s}-n_{1 + n,-s}}{(E_{1 + n,s}-E_{1 + n,-s})^2}+\frac{n_{n,s}-n_{2 + n,s}}{(E_{n,s}-E_{2 + n,s})^2}+\frac{n_{1 + n,-s}-n_{n,s}}{(E_{n,s}-E_{1 + n,-s})^2}\right)\times \\ 
&& \mathrm{Im} \left( \frac{\langle n,s|\delta M_x \sigma_x\tau_z|1 + n,s\rangle\langle 1 + n,s|J_i|2 + n,s\rangle\langle 2 + n,s|M_x\sigma_x|1 + n,-s\rangle\langle 1 + n,-s|\sigma_i|n,s\rangle}{(E_{1 + n,s}-E_{n,s}) (E_{2 + n,s}-E_{1 + n,-s})} \right) \\
&+& (M_x\sigma_x \leftrightarrow \delta M_x\sigma_x\tau_z)\Big].
\end{eqnarray*}
\end{widetext}
The terms before the last line correspond to $V_1 = \delta M_x\sigma_x\tau_z$ and $V_2 = M_x\sigma_x$, while the last line refers to terms with $V_1 = M_x\sigma_x$ and $V_2 = \delta M_x\sigma_x\tau_z$. 

\begin{figure}[ht!]
	\centering
	\includegraphics[scale=0.5]{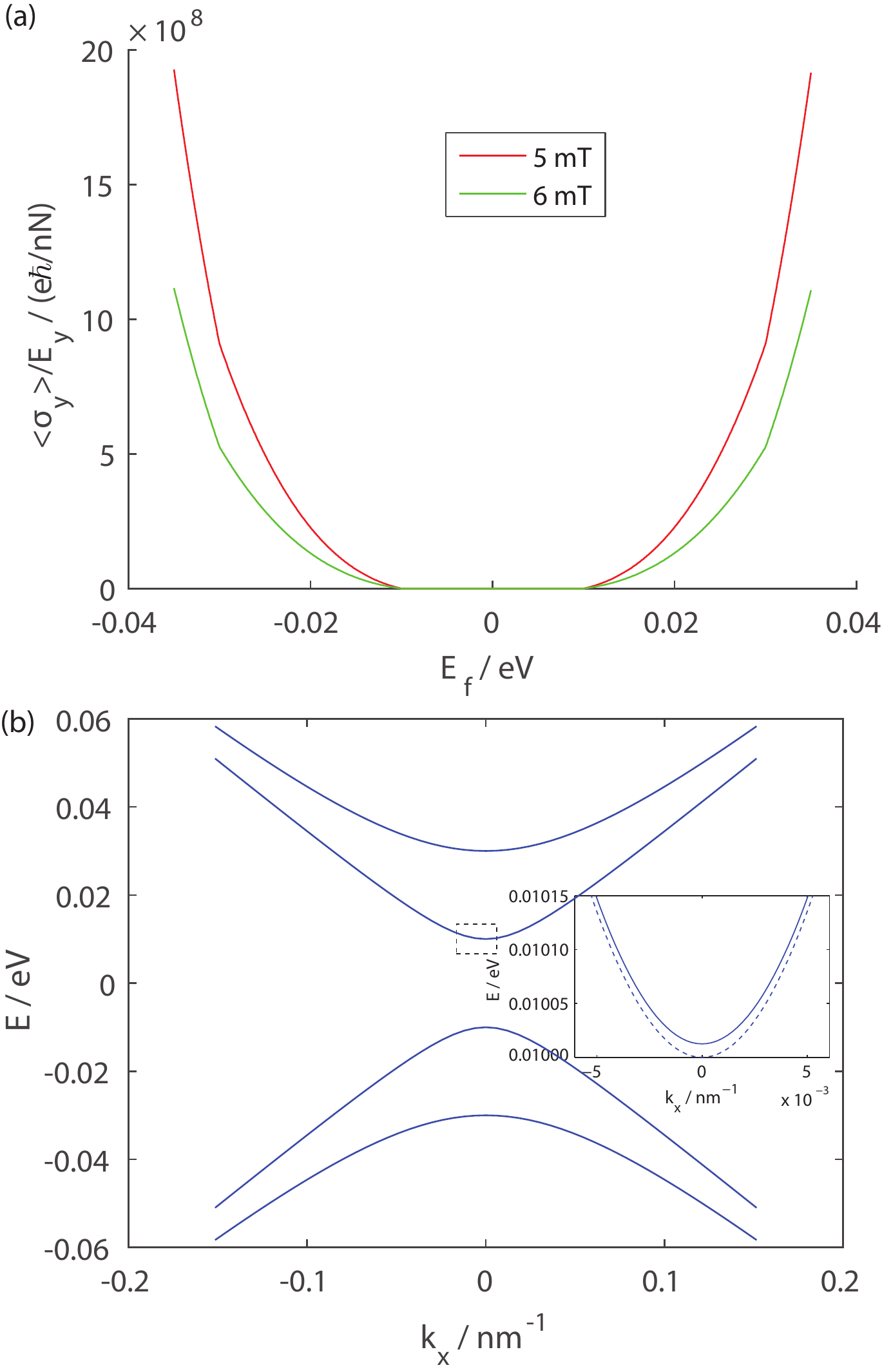}
	\caption{  (a) $\langle \sigma_y \rangle / E_x$ for $v_f = 5\times 10^5 m/s$ (which is typical for \ce{Bi2Se3}, $M_x = 5\ \mathrm{meV}$ $\Delta_t = 10\ \mathrm{meV}$ and $\Delta_z = 20\ \mathrm{meV}$ . (b) shows the dispersion relation of the TI thin film system \textit{without} the out of plane magnetic field at $k_y=0$. The dispersion relations between when the effective magnetization is $5\ \mathrm{meV}$ on the top and bottom surfaces, and when it is $5\ \mathrm{meV}$ on the top surface and $4.5\ \mathrm{meV}$ at the bottom one are not visually distinguishable at the scale of the plot. The inset shows a zoomed in view of the lowest energy particle band around $k_x=0$ showing the difference in the dispersion relation between when $\delta M_x$ has a finite value (solid line) and when $\delta M_x=0$ (dotted line). }
	\label{gA4b} 
\end{figure}	

Fig. \ref{gA4b}a shows the calculated $\langle \sigma_y \rangle / E_y$ for a TI thin film at various values of $B_z$ indicated on the figure legend \footnote{Since some of the authors are based in the engineering faculty our focus is on exploiting physical phenomena for potential device applications. The magnitudes of the magnetic fields we focus on here are closer to what may be technologically useful -- the fields are on the order of magnitude of that produced by a typical bar magnet -- rather than the multi-Tesla fields earlier works have focused on.} for an exemplary parameter set of $v_f = 5\times 10^5\ \mathrm{ms^{-1}}$, $\Delta_t = 10\ \mathrm{meV}$ and $\Delta_z = 20\ \mathrm{meV}$ and a magnetization of $5\ \mathrm{meV}$ in the $+x$ direction on the top surface, and $4.5 \mathrm{meV}$ on the bottom surface.      The  inset of panel (b) shows that the finite $\delta M_x$ leads to a small shift in the energies of the zero magnetic-field energy bands.  We study the spin accumulation in the $y$ direction as this is perpendicular to the magnetization direction and can exert a torque on the magnetization. 

The features present in the spin accumulation as a function of energy can be related to the dispersion relation in the absence of the out of plane magnetic field (panel (b)). We take this opportunity to discuss some features of the zero magnetic field band structure when $M_x \ll \Delta_t$, and $\delta M_x = E_z = 0$.  In this regime, the combination of the out of plane magnetization / Zeeman splitting $\Delta_z$ and in-plane magnetization $M_x$ leads to a bandgap of $2|\Delta_t - \sqrt{M_x^2 + \Delta_z^2}|$ between the particle-like and hole-like states. The inter-surface coupling lifts the degeneracies of the energy bands corresponding to states localized at the top / bottom surface, resulting in the formation of two particle (hole)-like bands where the energy increases (decreases) monotonically with $|k|$. The introduction of a small finite $\delta M_x$ (inset of panel (b) ) results in a shift in the energies of the bands.  

Referring back to Fig. \ref{gA4b} now, the zero spin accumulation Fermi energies at  $|E_f| < 10\ \mathrm{meV}$ corresponds to Fermi energies falling within the zero-magnetic field bandgap. The kinks in the spin accumulation near $E_f=0.03\ \mathrm{meV}$ in turn correspond to the emergence of the higher energy zero-field particle  subband. The increment of the spin accumulation with $E_f$ occurs at a slower rate above $E_f =0.03\ \mathrm{meV}$ because the contribution of the higher energy subband to the spin accumulation has an opposite sign to that of the lower energy particle subband. The magnitude of the spin accumulation increases with deceasing out of plane magnetic field.   (Despite the plot appearing to be symmetric about $E_f=0$ at the scale of the plot, this symmetry is in fact broken by the finite $\Delta_z$ \cite{PRB83_195413}. ) 

\section{Asymmetric potential} 
We now turn our attention to the effects of the $E_z\tau_z$ term. In the presence of a finite $E_z$, the Kronecker delta relations between $s$ and $s'$ in Eqs. \ref{Ez0Ji} and \ref{Ez0s}  no longer apply, so that now in general
\[
	\langle n\alpha s|O|n'\alpha' s'\rangle \propto \delta_{|n-n'|, 1},\ O=J_x,J_y,\sigma_x,\sigma_y
\]
has a finite value regardless of the relative signs of $s$ and $s'$. This implies that even in the absence of asymmetric magnetization ($\delta M_z=0$), there are diagrams up to second order in $M_x\sigma_x$ which give a finite $\langle \delta \sigma_i \rangle$ contribution. These diagrams are, namely, the zeroth order diagram which is Fig. \ref{gKuboDiag}a without the two $V$ vertices, and the second order diagram where the two vertices in Fig. \ref{gKuboDiag} both correspond to $M_x\sigma_x$. ( Similar to the $E_z=0$ case in the previous section, the first order diagrams which only has a $V=M_x\sigma_x$ vertex along the upper or the lower line do not give finite contributions because of the $n$ index mismatch. ) We do not consider the zeroth order diagram here, as it does not capture the in-plane magnetization. The explicit expression for the second order $\langle \delta \sigma_y \rangle$ contribution (not shown) is almost as cumbersome as the corresponding expression shown earlier for finite $\delta M_x$, and has a similar form except that we now need to sum over all the internal $s$ indices as well. 

\begin{figure}[ht!]
	\centering
	\includegraphics[scale=0.5]{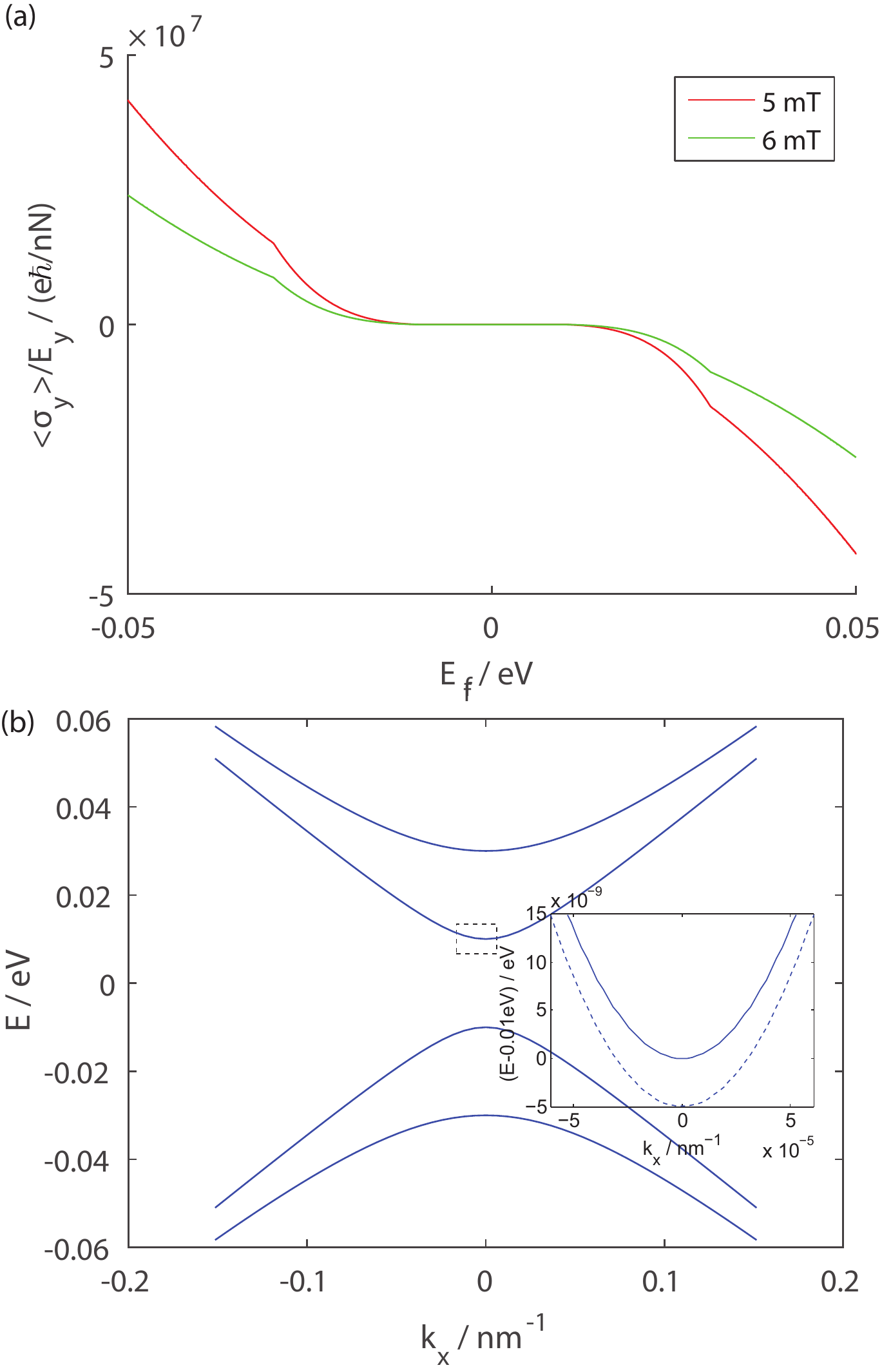}
	\caption{ (a) $\langle \sigma_y \rangle / E_x$ for a $v_f = 5\times 10^5 \mathrm{m/s}$ , $\Delta_t = 10\ \mathrm{meV}$ and $\Delta_z = 20\ \mathrm{meV}$, $M_x = 5\ \mathrm{meV}$, $\delta M_x = 0$, $E_z = 0.01\ \mathrm{meV}$ TI thin film system due to an electric field in the $y$ direction.   (b) shows the dispersion relation of the TI thin film system \textit{without} the out of plane magnetic field at $k_y=0$. The dispersion relations when $E_z$ is finite, and when it is 0,  is not visually distinguishable at the scale of the plot.  The inset shows a zoomed in view of the lowest energy particle band around $k_x=0$ showing the difference in the dispersion relation between when $E_z$ has a finite value (solid line) and when $E_z=0$ (dotted line.) } 
	\label{gA2b} 
\end{figure}

Fig. \ref{gA2b} shows the spin $y$ accumulation due to an electric field in the $x$ direction for the same set as parameters as in Fig. \ref{gA4b} with the exceptions that here $M_x  = 5\ \mathrm{meV}$ for both the top and bottom surfaces, and $E_z = 0.01\ \mathrm{meV}$.  

Similar to the case where there is asymmetric magnetization, the second order $M_x$ spin accumulation due to $E_y$ is zero when the Fermi energy falls within the zero-field band gap and increases with decreasing out of plane field. The rate of increase of the spin accumulation with $E_f$ here also decreases once $E_f$ rises above the band bottom of the higher energy zero field band because the contribution of this band to the spin accumulation is of opposite sign to that due to the lower energy particle band. Differing from the contribution due to asymmetric magnetization, the sign of the spin accumulation here switches with the sign of $E_f$. 

\section{Origin of spin accumulation} 
\begin{figure}[ht!]
	\centering
	\includegraphics[scale=0.4]{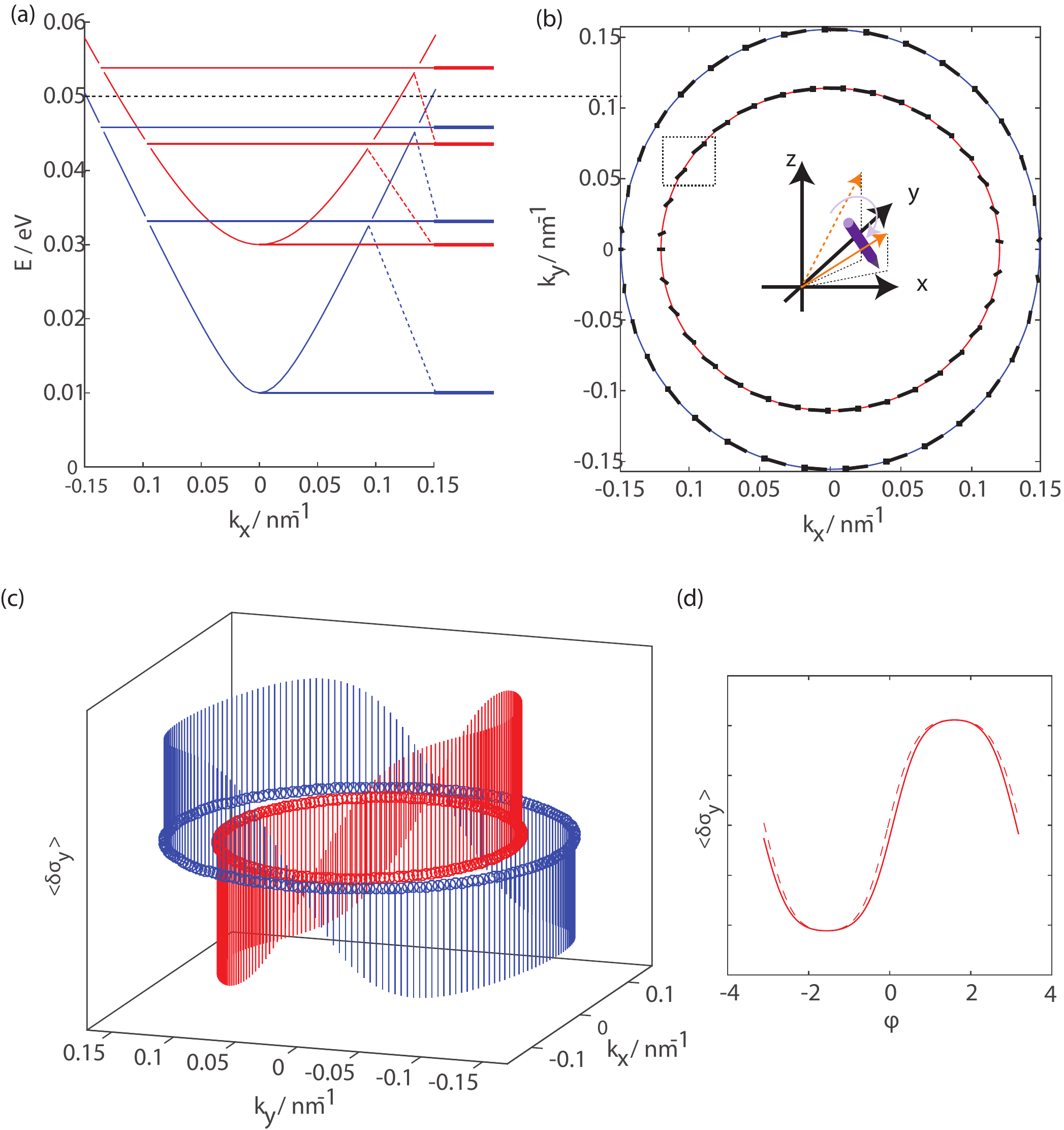}
	\caption{(a)  The collapse of the zero-field energy bands for the parameters in Figs. \ref{gA2b} and \ref{gA4b} (except that $E_z=\delta M_x=0$) into the $s=1$ (red) and $s=-1$ (blue) Landau levels for $B=60\ \mathrm{T} $, represented by the thick lines on the right. (b) shows the directions and relative magnitudes of the in-plane spin accumulation of the two zero-field bands at $E=0.05\ \mathrm{eV}$. The inset shows how the rotation of the spins (from the direction of the dotted green arrow to the solid green arrow) at a given $\vec{k}$ value as the $k_y$ value is shifted upwards implies the existence of an effective exchange field in order to exert the torque (solid purple arrow) needed to rotate the field, and which confers a corresponding spin accumulation in the direction of the field. (c) shows the relative magnitudes of the spin $y$ accumulation at different points on the $E=0.05\ \mathrm{eV}$ zero-field equal energy contour. Panel (d) shows a finite $\delta M_x$ breaks the exact antisymmetry of the spin accumulation. The solid line is the spin $y$ accumulation plotted as a function of angular coordinate $\phi$ around the EEC of the higher energy particle band (i.e. the smaller circle) for a finite $M_x$. The dotted line is the solid line reflected about both $\phi=0$ and  $\langle \delta \sigma_y \rangle=0$. The fact that the line and its reflection do not overlap exactly indicates that spin $y$ accumulation is not exactly antisymmetric. } 
	\label{gdelSyKy} 
\end{figure}

We offer an intuitive explanation of the spin accumulation. Each discrete Landau level may roughly be thought of as coming from the collapse of the zero magnetic field states in the energy vicinity of the Landau level into a single value of energy, as illustrated schematically in Fig. \ref{gdelSyKy}(a). (We have exaggerated the values of $B$ and $\delta M_x$ in the figure compared to the parameters in our actual calculations for expositional ease -- the features we highlight would not have been visible at the scale of the figures otherwise. ) . 
 
In our system, each of the two zero-field subbands collapses into the Landau level states $|n \alpha s\rangle$ with different  $s$ indices.  One may therefore gain some insights about  the spin accumulation in a Landau level by studying the spin accumulation in the constituent zero-field states that make up the Landau level.  Panel(b) of the figure shows the in-plane spin accumulation directions $\langle \vec{\sigma} \rangle$ along the $E = 50\ \mathrm{meV}$ equal energy contours for the two particle bands present at the parameter set of Figs. \ref{gA4b} and \ref{gA2b} with $\delta M_x = E_z = 0$. 

The spin accumulation at each point on the EEC may be thought of as being due to a $\vec{k}$-dependent spin-orbit interaction field $\vec{b}(\vec{k})$. Applying an electric field in the $y$ direction causes an small shift in the $k_y$ component of each point on the EEC so that the spin at each $\vec{k}$ point now adiabatically rotates to point to the direction of the spin-orbit interaction field at the new, $k_y$ shifted $\vec{k}$ point (inset of panel(b)).  ( Refer to Ref. \onlinecite{Ar1606_03812} for more details.) The electric field induced rotation of the spin accumulation at each $\vec{k}$ point may be thought of as being due to an effective exchange field pointing in the $\hat{b}\times\partial_{k_b}\hat{b}$ direction  which not only provides the torque needed to effect the rotation but also confers a spin accumulation in the direction of the effective exchange field \cite{Ar1606_03812,Grp1,Grp2,Grp3,Grp4,SGTSciRep}. Panel (c) of the figure shows the distribution of the resulting $E_y$ induced spin $y$ accumulation on each of the EEC points. 

In the absence of the $E_z\tau_z$ and $\delta M_x\sigma_x\tau_z$ terms the spin $y$ accumulation is antisymmetric and cancels out exactly. The introduction of either term breaks the exact antisymmetry (panel (d)) of the spin accumulation and results in a finite spin $y$ accumulation after summing over the entire EEC, and a further sum over the energy ranges falling within a Landau level after the out of plane magnetic field is applied.  

\section{Conclusion} 
In this work we studied the electric-field induced spin accumulation in a topological insulator thin film system with an out of plane magnetic field and an in-plane magnetization. We showed that the second order perturbative calculation for the in-plane magnetization adequately reproduces the exact energy spectrum, and then used the Kubo formalism to calculate the spin accumulation perpendicular to the magnet due to an electric field. The electric field does not lead to a finite spin accumulation to second order in the magnetization in a TI thin film which has inversion symmetry with respect to the top and bottom surfaces due to the restrictive relations linking the Landau level matrix elements of the magnetization, spin and current operators involved in the Kubo calculation. 
We then saw that the introduction of two types of asymmetry  -- (a) a scalar potential difference and (b) differing magnetization magnitudes -- between the top and bottom surfaces of the film -- relaxes these restrictions and leads to the emergence of finite spin accumulation for Fermi energies falling outside the zero-magnetic field bandgap. This spin accumulation results from the breaking of the antisymmetry of the spin accumulation around the zero-magnetic field equal energy contours.

\section{Acknowledgments} 
ZBS and MBAJ acknowledge the Singapore National Research Foundation for support under NRF Award Nos. NRF-CRP9-2011-01 and NRF-CRP12-2013-01, and MOE under Grant No. R263000B10112.

 
\end{document}